\documentstyle[11pt,aaspp4]{article}
\input{psfig}
\eqsecnum
\def\beq{\begin{equation}}
\def\eeq{\end{equation}}
\def\ref{\reference}
\def\simge{\mathrel{%
   \rlap{\raise 0.511ex \hbox{$>$}}{\lower 0.511ex \hbox{$\sim$}}}}
\def\simle{\mathrel{
   \rlap{\raise 0.511ex \hbox{$<$}}{\lower 0.511ex \hbox{$\sim$}}}}

\begin{document}
\title{Spin-Up of Low Luminosity Low Mass X-ray Binaries}
\author{Insu Yi$^1$ and Jonathan E. Grindlay$^2$}
\affil{$^1$Institute for Advanced Study, Princeton, NJ 08540}
\affil{$^2$Harvard-Smithsonian Center for Astrophysics, 60 Garden St., 
Cambridge, MA 02138}

\begin{abstract}

We examine the spin-up of low luminosity, low mass X-ray binaries (LMXBs) 
to millisecond pulsars (MSPs). In the conventional spin-up model of the 
Ghosh \& Lamb type, where the stellar magnetic field interacts with the 
Keplerian accretion disk, MSPs could be produced from LMXBs if magnetic 
field $B_*\simle 10^{8}({\dot M}/10^{16}g/s)^{1/2}G$, where ${\dot M}$ 
is the mass accretion rate. However, for ${\dot M}<{\dot M}_c\sim 10^{16}g/s$ 
accretion is likely to occur via a quasi-spherical flow with a sub-Keplerian 
rotation. The sub-Keplerian rotation rate is smaller than the Keplerian rate 
by a factor $\sim 2-10$. As a consequence, the spin-up of LMXBs produces 
pulsars with spin periods longer by a factor $\sim 2-10$ than those with a 
Keplerian accretion disk. The observed MSPs could be produced only for 
$B_*<10^7G$ even when ${\dot M}\sim {\dot M}_c\sim 10^{16}g/s$. 
This suggests that the low luminosity LMXBs with ${\dot M}<{\dot M}_c$ 
would not be able to spin-up to the observed MSPs. This rules out any 
undetected populations of persistent, low luminosity LMXBs and potentially 
a significant fraction of the atoll sources as a possible source population 
of the observed MSPs. If a large number of undetected, persistent, low 
luminosity LMXBs do exist, they could produce MSPs near the pulsar death line 
with intrinsic electromagnetic luminosity $\simle 10^{30} erg/s$. 
The observed MSPs could possibly arise from a population of 
soft x-ray transients (SXTs) containing neutron stars although this 
is not supported by current estimates of the outburst rate or numbers
of such systems. Accretion induced collapse of low magnetic field
white dwarfs remains a possible channel for MSP formation. 
\end{abstract}

\keywords{accretion, accretion disks $-$ pulsars: general
$-$ stars: magnetic fields $-$ X-rays: stars}

\section{Introduction}

The origin of millisecond pulsars (MSPs) remains an outstanding problem
(e.g. Bhattacharya \& van den Heuvel 1991). A number of channels of MSP 
formation have been discussed. In the standard model, magnetized,
low mass X-ray binaries (LMXBs) spin-up to MSPs through mass accretion
(e.g. Frank et al. 1992). Several statistical analyses, however, have 
questioned whether the known LMXB populations could produce the observed MSPs
(Grindlay \& Bailyn 1988, Kulkarni \& Narayan 1988, Bailyn \& Grindlay 1990, 
Kulkarni 1995).

The problem of MSP birth rate is briefly summarized as follows (e.g. Grindlay 
1995). The MSP lifetime due to the electromagnetic dipole emission (e.g. 
Shapiro \& Teukolsky 1983) is 
$\tau_{MSP}\sim 10^{10}(B_*/5\times 10^8G)^{-2}(P_{*}/5ms)^2~yr$ where $B_*$ 
is the pulsar magnetic field strength and $P_*$ is the spin period. For the 
estimated total number of MSPs (Bailes \& Lorimer 1995), $N_{MSP}\sim 10^5$, 
the observed MSPs require a birth rate of $\sim N_{MSP}/\tau_{MSP}\sim 10^{-5}$
on the MSP birth rate, $\sim 3\times 10^{-6}~yr^{-1}$ above the luminosity
limit of 1 mJy kpc$^2$.
The uncertainty in this estimate could be large due to uncertainties 
in beaming of pulsar emission and the Galactic scale height of the pulsar 
distribution. For a high accretion rate ${\dot M}$ and a relatively low $B_*$, 
the accretion flow extends close to the neutron star surface. Then, 
the spin-up of a slowly rotating neutron star to the spin period of
$P_*\sim 5ms$, would require mass accretion of at least
$\simge \Delta M\sim I_*\Omega_*/(GM_*R_*)^{1/2}\sim 0.1M_{\odot}$, 
where $I_*=10^{45} gcm^2$ is the moment of inertia, $M_*=1.4M_{\odot}$ 
is the mass, $R_*=10^6cm$ is the radius of the neutron
star, and $\Omega_*=2\pi/P_*$, respectively (cf. Bhattacharya \& van den Heuvel
1991). Luminous Z sources have luminosities 
$L_Z\sim 3\times 10^{37}-10^{38} erg/s$ close to the Eddington luminosity 
$L_{Edd}\sim 2\times 10^{38} erg/s$ and their magnetic fields are  
estimated to be in the range $\sim 10^{9-10}G$.
The typical spin-up time scale for Z sources 
$\tau_Z\simge GM_*\Delta M/L_ZR_*\sim 3\times 10^7~yr$.
For the observed number of Z sources $N_Z=6$, the MSP birth rate from Z sources 
is expected to be $\simle N_Z/\tau_z\sim 2\times 10^{-7} yr^{-1}$. 
For less luminous, bursting atoll sources with luminosities
$L_{atoll}\sim 10^{36}-10^{37} erg/s$, the typical spin-up time scale
$\tau_{atoll}\simge GM_*\Delta M/L_{atoll}R_*\sim 10^9 yr$. For the total 
number of atoll sources $N_{atoll}\sim 10^2$, the estimated birth rate is
$\simle N_{atoll}/\tau_{atoll}\sim 10^{-7} yr^{-1}$, which is within a factor
$\sim 2$ of the Z source rate. These rates fall short of the MSP formation rate
roughly by at least an order of magnitude (cf. Kulkarni 1995). 

Recent observations with RXTE (cf. summary in White and Zhang 1997) have 
yielded the long-sought evidence that LMXBs (including both Z sources and 
burst sources) are spun-up to $\sim$few ms spin periods as indicated 
by their kHz QPOs and the direct detection (in a few sources) of what appear 
to be underlying spin periods. Some 10 sources now appear to have $\sim300Hz$ 
pulsation frequencies which certainly suggests that these sources have indeed 
spun-up and could thus evolve to MSPs. 
Recently, the first direct detection of a true LMXB-MSP 
(SAXJ1808.4-3658) with a spin-period of 2.49ms has been 
reported (Wijnands and van der Klis 1998) along with the 
value for its binary period ($\sim$2h), probable companion mass 
($\simle0.1M_{\odot}$) and thus likely evolution (Chakrabarty 
and Morgan 1998). However, 
even if all the known LMXBs are found 
to have similar pulsation periods, the birth rate problem summarized 
above still remains to be resolved.

The MSP formation problem could be solved if 
a generally unseen population of LMXBs with low luminosities
spin-up to MSPs. Recent observations have measured low quiescent luminosities 
from recurrent transients as well as some usually bright persistent sources 
(Verbunt et al. 1994). The required number $N_l$ of the low 
luminosity LMXBs with the luminosity $L_l$ is 
$N_l\sim 10N_{atoll}(L_{atoll}/L_l)$ if these sources are the solution to the 
MSP formation problem. Since the spin-up time scale for the low luminosity
population $\tau_l\sim \tau_{atoll}(L_{atoll}/L_l)$, $L_l$ must be at least
$\sim 0.1L_{atoll}$ in order to avoid an excessively long spin-up time scale.
$N_l\sim 100N_{atoll}$ is required with $L_l\sim 0.1L_{atoll}$. The low
luminosities are either a constant for persistent sources or a time-averaged 
value for recurrent transient sources. 
It is crucial whether low luminosity sources with 
$L_l\sim 0.1L_{atoll}$, persistent or recurrent, are able to spin up to MSPs. 

\section{Spin-Up of LMXBs: Standard Model with Low ${\dot M}$}

The spin period of the neutron star evolves according to
${\dot P}_*=-NP_*^2/I_*$,
where $N$ is the torque exerted on the star by the accretion flow.
In the conventional magnetized, Keplerian accretion disk mode (e.g. Wang
1995, Yi et al. 1997, and references therein),
the accretion flow is truncated by the neutron star magnetosphere at a
radius $R_o$ determined by
\beq
(R_o/R_c)^{7/2}\left[1-(R_o/R_c)^{3/2}\right]^{-1}=2N_c/{\dot M}(GM_*R_c)^{1/2}
\eeq
where $R_c=(GM_*P_*^2/4\pi^2)^{1/3}$ is the Keplerian corotation radius,
$N_c=(\gamma/\alpha)B_*^2R_*^6R_c^{-3}$, $\alpha$ is the usual viscosity 
parameter (Frank et al. 1992), and $\gamma$ is the parameter of order unity 
which determines the pitch of the magnetic field in the accretion disk
(e.g. Wang 1995).
For numerical values, we adopt $\alpha=0.3$ and $\gamma=1$ unless noted
otherwise. The torque exerted on the star
\beq
N=(7N_o/6)\left[1-(8/7)(R_o/R_c)^{3/2}\right]/\left[1-(R_o/R_c)^{3/2}\right]
\eeq
where $N_o={\dot M}(GM_*R_o)^{1/2}$.
The spin equilibrium $N=0$ occurs when $R_o/R_c=x_{eq}=(7/8)^{2/3}$.
This adopted magnetic torque model differs little from other phenomenological 
models for the purpose of the present discussion (Wang 1995).

In the spin-up regime $N>0$ or $R_o/R_c<x_{eq}$, we get from eq. (2-2)
\beq
N\approx 1.3(GM_*R_*^2)^{3/7}(\gamma/\alpha)^{1/7}B_*^{2/7}{\dot M}^{6/7}.
\eeq
In this regime, the spin period evolves simply as ${\dot P}_*\propto
-(\gamma/\alpha)^{1/7}B_*^{2/7}{\dot M}^{6/7}P_*^2$.
The spin-up line is determined by the equilibrium spin condition
$N=0$,
\beq
P_*\approx [5ms]\gamma^{3/7}\alpha_{-1}^{-3/7}B_{*,8}^{6/7}{\dot M}_{16}^{-3/7}
\eeq
where $\alpha_{-1}=\alpha/0.1$, $B_{*,8}=B_*/10^8G$, and 
${\dot M}_{16}={\dot M}/10^{16} g/s$. Spin periods of
$P_*\simle 5ms$ would be reached for
$B_*<[3\times 10^8 G](\alpha/\gamma)^{1/2}{\dot M}_{16}^{1/2}$.
The spin-up line corresponding to ${\dot M}=10^{18} g/s$, which
is close to the Eddington rate, is $P_*\approx [0.4ms]B_{*,8}^{6/7}$,
which obviously implies that MSPs with $P_*\sim 5ms$
could easily form from high luminosity LMXBs such as Z sources 
with $B_*$ as high as $\simle 3\times 10^9G$. For ${\dot M}=10^{16} g/s$, 
which is relevant for atoll sources, the spin-up line is
$P_*\approx [3ms]B_{*,8}^{6/7}$.
$P_*\sim 5ms$ requires $B_*\simle 2\times 10^8G$. Given the 
dependence of $B_*\propto (\alpha/\gamma)^{1/2}$, it is gratifying that
the uncertainties in $\alpha$ and $\gamma$ are not critical unless
$\alpha\ll 0.1$ and $\gamma\gg 1$, which is highly unlikely 
(e.g. Aly \& Kuijpers 1990, Wang 1995).

In Figure 1, we plot the spin-up lines in the $P_*$ vs. $B_*$ plane. 
The x's mark the positions of the observed pulsars with
their spin periods and estimated magnetic fields adopted from Taylor
et al. (1995). The magnetic fields have been estimated based on the 
assumption that their power is derived from the electromagnetic dipole 
emission at a rate ${\dot E}=4\pi I_*{\dot P}_*/P_*^3$; i.e.
$B_*=(3I_*c^3P_*{\dot P}/8\pi^2R_*^6)^{1/2}$.
In Figure 1a, we show spin-up lines assuming that the accretion flow
remains Keplerian for all ${\dot M}$'s. This plot
shows that MSPs form only from LMXBs with weak magnetic fields 
$<10^8 G$ if ${\dot M}<6\times 10^{15}~g/s$ or LMXBs with 
$B_*<10^7 G$ if ${\dot M}\sim 10^{13}~g/s$. 
A system with a lower ${\dot M}$ needs proportionally more time to spin-up to 
a MSP than does a high ${\dot M}$ system. The final spin period becomes
the equilibrium period, which is fixed by both ${\dot M}$ and $B_*$, only if
the system arrives at the equilibrium period on timescales for the binary 
not to have evolved significantly. Otherwise, ${\dot M}$ would have changed 
significantly, in which case a new equilibrium spin period would be sought.
For ${\dot M}\sim 10^{13} g/s$, the spin-up time scale is prohibitively 
long $\gg 10^{10} yr$ and spin-up to a MSP in a time-scale
less than the age of the Galaxy is impossible.
Therefore, in the standard model, the MSP formation rate problem could be
solved only if there are $\simge 5\times 10^2N_{atoll}\sim 5\times 10^4$ 
persistent, low luminosity LMXBs with ${\dot M}\sim 10^{15}~g/s$ (i.e. 
luminosity $L\sim 5\times 10^{-2}L_{atoll}$) and magnetic fields 
$\sim 5\times 10^{7}G$ (see below for recurrent sources).

\section{Spin-Up by Sub-Keplerian Flows at Low Mass Accretion Rates}

The conclusion we have derived above may get dramatically worse due to the
possible transition in accretion flows (Grindlay 1995).
This is because at low ${\dot M}$, the accretion flow is likely to take the 
form of a  
quasi-spherical, sub-Keplerian rotation (e.g. Narayan \& Yi 1995).
More specifically, when ${\dot M}$ falls
below a critical accretion rate ${\dot M}_c$, the viscously dissipated energy
in the accretion flow is not efficiently radiated but kept as internal heat 
energy of the accreted gas.
As a result, the accretion flow is heated essentially to the virial temperature
and the accretion flow thickens due to the internal pressure support. 
The internal pressure support also causes the rotation of the quasi-spherical 
flow to deviate from Keplerian to sub-Keplerian.

The sub-Keplerian rotation of the accretion flow
can be conveniently modeled as $\Omega=\omega_a\Omega_K$ where $\omega_a\le 1$ 
where $\Omega_K=(GM_*/R^3)^{1/2}$ is the Keplerian rotation 
(Narayan \& Yi 1995, Yi et al. 1997).
The sub-Keplerian rotation parameter $\omega_a$ is determined 
essentially by the ratio of the internal pressures 
$\beta=P_{gas}/(P_{gas}+P_{mag})$ 
where $P_{gas}$ is the gas pressure and $P_{mag}$ is the
magnetic pressure from the tangled, isotropic magnetic field in equipartition
with the thermal energy of the accreted plasma; i.e.
$\omega_a=[(10-6\gamma_s)/(9\gamma_s-5)]^{1/2}$ where 
$\gamma_s=(8-3\beta)/(6-3\beta)$ is the ratio of specific heats in the assumed 
equipartition plasma (Esin 1997, cf. Narayan \& Yi 1995). 
A large $\beta\sim 1$ (i.e. small magnetic pressure) would give
$\omega_a\ll 1$ whereas $\omega_a\sim 0.4$ for equipartition $\beta\sim 0.5$.
The sub-Keplerian rotation parameter $\omega_a$ weakly depends on $\alpha$
as long as $\alpha=0.01-0.3$.
Recently, Yi et al. (1997) have suggested that the torque reversals seen in 
some X-ray pulsars could be precisely due to this accretion flow 
transition, which would also confirm that indeed sub-Keplerian flows 
form probably for ${\dot M}$ below a critical rate ${\dot M}_c$ for accretion
onto NSs since otherwise these advection dominated accretion flows (ADAFs) have 
been derived originally (Narayan and Yi 1995)  only for accretion onto BHs.  
It has been demonstrated that observed reversal events are reproduced by 
$\omega_a\approx 0.1-0.4$ when the transition occurs at 
${\dot M}={\dot M}_c\sim 10^{16} g/s$. 
This supports the existence of the plasma roughly with
equipartition in accretion flows around LMXBs when ${\dot M}<{\dot M}_c$.

The critical accretion rate ${\dot M}_c\sim 0.1\alpha^2{\dot M}_{Edd}$ 
where ${\dot M}_{Edd}$ is the Eddington accretion rate with 10\% efficiency
(Narayan \& Yi 1995). For $\alpha=0.3$, ${\dot M}_c\sim 10^{16}g/s$ is
expected. 
This rate could be somewhat lower than the 
critical rate for the ADAF black hole systems, since 
cooling of the accretion flow is more efficient 
in neutron star systems due to the soft photons from the neutron star surface 
(Narayan \& Yi 1995). 
The appearance of the quasi-spherical flow is accompanied
by slight luminosity changes (Yi et al. 1997 and references therein) but the
X-ray spectral changes appear to be significant (Vaughan \& Kitamoto 1997).

When the accretion flow rotation becomes sub-Keplerian, 
the corotation radius is shifted to a new location,
$R_c^{\prime}=\omega_a^{2/3}R_c$ (Yi et al. 1997). 
The magnetospheric radius $R_o$ is then shifted to the radius $R_o^{\prime}$ 
which is determined by (cf. eq. 2-1)
\beq
(R_o^{\prime}/R_c^{\prime})^3\left[1-(R_o^{\prime}/R_c^{\prime})^{3/2}\right]
^{-1}=2N_c/\omega_a N_o^{\prime}
\eeq
where $N_o^{\prime}=\omega_a{\dot M}(GM_*R_o^{\prime})^{1/2}$.
The torque exerted on the star is modified as (Yi et al. 1997)
\beq
N^{\prime}=(7N_o^{\prime}/6)\left[1-(8/7)(R_o^{\prime}/R_c^{\prime})^{3/2}
\right]/\left[1-(R_o^{\prime}/R_c^{\prime})^{3/2}\right].
\eeq
In the spin-up regime, the spin-up torque is given by
\beq
N^{\prime}\approx 1.3\omega_a(GM_*)^{3/7}R_*^{6/7}(\gamma/\alpha)^{1/7}
B_*^{2/7}{\dot M}^{6/7},
\eeq
and the spin-up line is given by
\beq
P_*=[49ms]
\omega_{a,-1}\gamma^{3/7}\alpha_{-1}^{-3/7}B_{*,8}^{6/7}{\dot M}_{16}^{-3/7}
\eeq
where $\omega_{a,-1}=\omega_a/0.1$. The equilibrium spin period for given 
${\dot M}$ and $B_*$ becomes longer by a factor $\omega_a^{-1}>1$.
$P_*\simle 5ms$ would be reached for
\beq
B_*\simle [3\times 10^8G]\omega_a^{7/6}(\alpha/\gamma)^{1/2}{\dot M}_{16}.
\eeq
For $\omega_a=0.2$ and ${\dot M}=10^{16} g/s$, $P_*\approx [15ms]B_{*,8}^{6/7}$,
which suggests that a typical MSP with $P_*=5ms$ would require
$B_*\simle 3\times 10^7G$. Any LMXBs with ${\dot M}<10^{15} g/s$ cannot
produce the observed MSPs with $B_*> 5\times 10^7G$.

It is usually thought that neutron stars are born with larger 
magnetic fields and that such low magnetic fields as required 
($B_*\simle 3\times 10^7G$) for 
spinup of low luminosity LMXBs to MSPs must therefore result from 
field decay. Although the exact mechanism for the 
field decay still remains uncertain, in a simple heuristic model, the field 
decays as a result of mass accretion and the strength of the magnetic field 
is determined by the amount of mass accreted (e.g. Taam \& van den Heuvel 
1986, Romani 1990). 
In this model, $B_*\propto ({\dot M}\Delta t/\Delta M_{d})^{-x}$,
where ${\Delta t}$ is duration of accretion and ${\Delta M}_d$ is the
characteristic mass scale for field decay. Assuming that 
neutron stars are typically produced with 
field strengths $\simge 10^{12}G$, the spin-up of atoll 
sources with ${\dot M}\sim 10^{17}~g/s$ to MSPs requires
$\Delta t\sim 5\times 10^{7}yr$ and $B_*\sim 3\times 10^8G$ 
(and $\sim 6\times 10^{8}$ yr 
for the sub-critical accretion case with 
${\dot M}\sim 10^{16}$~g/s and $B_*\simle 3\times 10^7G$). 
Thus for the atoll source case (with  ${\dot M}\sim 10^{17}~g/s$),  
the field decay model requires that the characteristic mass 
$\Delta M_d\sim 3\times 10^{-5}M_{\odot}$ for $x\sim 1$. 

However, there is now evidence against field decay (in LMXBs) 
to values below $\sim$10$^8$ G: the  
newly discovered LMXB-MSP SAXJ1808.4-3658 (Wijnands 
and van der Klis 1998) directly shows this LMXB contains a NS 
with a likely magnetic field B $\sim (2-14)\times 10^8$ G (although 
these same authors speculate that perhaps the lack of detection of 
pulsations in other LMXBs suggests they may indeed have lower B
fields). Additional evidence against field decay (to below 10$^8$ G) 
in even very old LMXB systems comes from the detection of 
kHz QPOs and possible pulsations (at the QPO difference frequency 
of 275$\pm$8 Hz) in the relatively luminous atoll source 4U1820-30 in the 
globular cluster NGC 6624 (Smale et al. 1997, Zhang et al 1998). 
Since the NS in this 
system is a Pop II NS, and therefore likely to have an age $\sim$10$^{10}~yr$, 
and yet the kHz QPOs (and possible pulsations) in this object 
with ${\dot M}\approx 10^{17} g/s$ suggest a magnetic field
$B\simge 3\times 10^8$ G, it appears that even continued accretion has not 
reduced the B field below this value. 

Thus the required B field values of MSPs actually observed,  
the required spin-up timescales, and the lack of evidence 
that field decay continues below  $\sim$10$^8$ G combine to 
effectively rule out any persistent low luminosity sources as a possible
source of MSPs. In fact, spin-up of atoll sources to MSPs also becomes
questionable. This rather radical conclusion could be 
avoided only when $\alpha/\gamma\gg 1$ or when ${\dot M}_c\ll 10^{16} g/s$. 
The former possibility is not likely (Wang 1995, Aly \& Kuijpers 1990) and
the latter is not supported by the recent work 
(e.g. Narayan \& Yi 1995, Yi et al. 1997, and references therein).

In Figure 1b, we show the spin-up lines assuming that
the accretion flow becomes sub-Keplerian below the assumed critical accretion
rate $3.5\times 10^{16}g/s$ with $\omega_a=0.2$ (Yi et al. 1997).
This figure confirms our conclusion that even at a relatively high accretion 
rate $\sim 10^{16} g/s$, MSPs would be produced only when 
$B_*\simle 3\times 10^{7}G$. 
For such low $B_*$'s, most of LMXBs with ${\dot M}<6\times 10^{15}~g/s$ 
would not emerge out of the so-called pulsar graveyard below the death line
(e.g. Ritchings 1976) during spin-up. They would be detected as MSPs only 
after arriving at periods $P_*\simle 3\times 10^{-2} s$ (i.e. solid lines 
above the the death line). It is interesting to point out that 
in the $P_{*}$ vs. $B_*$ plot, there could be a gap due to the accretion
flow transition as implied by the spin-up line for the assumed
${\dot M}_c\sim 3.5\times 10^{16}g/s$. 
For ${\dot M}>{\dot M}_c$ the Keplerian spin-up line 
in Figure 1a applies whereas  for ${\dot M}<{\dot M}_c$ (Figure 1b)
the sub-Keplerian spin-up 
line would be shifted discontinuously downward, leaving a region between the 
two lines as a gap. If there is a significant population of low luminosity 
LMXBs spun-up with the sub-Keplerian rotation, they could produce MSPs above
the death line typically with $P_*\sim 5ms$ and $B_*\sim 10^7G$. These MSPs
would populate the region marked by the solid box in the bottom left corner 
in Figure 1b. Their intrinsic electromagnetic power is expected to be only
$\sim B_*^2R_*^6\Omega_*^4/6c^3\simle 2\times 10^{30} erg/s$, which is roughly
two orders of magnitude lower than that of the typical MSPs 
and is therefore likely well below the $\sim$1mJy-kpc$^2$ radio 
luminosity limit of MSPs detected by Lyne et al (1997).
It is not clear whether a significant number of such MSPs could be found 
in reality given the longer spin-up time scales ($\simge 10^9yr$) for lower 
${\dot M}$'s.

\section{MSPs from Reccurent Soft X-ray Transients}

Some LMXBs show low luminosities 
(i.e. ${\dot M}<{\dot M}_c$)  between
recurrent outbursts (Verbunt et al. 1994) or during some periods as
persistent sources (e.g. Tanaka \& Shibazaki 1996). During high ${\dot M}$, 
the spin-up time scale and the equilibrium spin period could become considerably
shorter than those of the low ${\dot M}$ period. If the high ${\dot M}$ and low
${\dot M}$ periods alternate, the LMXB systems would be constantly away from 
their equilibrium spin periods. The typical spin-up time scale 
$t_{spin-up}=P_*/{\dot P}_*$ depends on ${\dot M}$ as
$$
t_{spin-up}=I_*/P_*N\approx 2(M_*R_*^2)^{4/7}/(\gamma/\alpha)^{1/7}
(G{\dot M}^2)^{3/7}B_*^{2/7}P_*
$$
\beq
\approx [5\times 10^7 yr](\gamma/\alpha/10)^{1/7}
{\dot M}_{16}^{-6/7}B_{*,8}^{-2/7}P_{*,1}^{-1}
\eeq
where $P_{*,1}=P_*/10ms$. This time scale applies to both the Keplerian 
accretion disk and the sub-Keplerian flow.
If recurrent outbursts have ${\dot M}$ as high as $\sim 10^{18} g/s$,
the short equilibrium spin period corresponding to this high ${\dot M}$ 
would be reached if the outburst phase cumulatively lasts for $\simge 10^7yr$ 
while accreting a mass of $\sim 0.1M_{\odot}$. 
Since the low ${\dot M}<10^{15} g/s$ phase would have a long spin-up time 
scale $>10^9 yr$, the recurrent outbursts 
would determine the eventual spin period if the duration of the outburst is
$\simle 10^{-3}$ of the quiescent duration. For instance, for an outburst
recurrence time scale $\sim 10yr$, the required duration of outbursts is
$\sim 1 month$, which is typical of the 
``soft x-ray transients" (SXTs) containing black hole primaries 
(e.g. Tanaka \& Shibazaki 1996). 

SXT systems containing neutron star primaries (NS-SXTs) could thus in 
principle solve the MSP formation problem. 
The plausibility of this channel of MSP formation
critically depends on actual detections of NS-SXTs in sufficient
numbers. Presently, there is little evidence which indicates the existence of 
a large population of such transient sources, and in fact van Paradijs (1996) 
has shown that among the transients BH-SXTs are likely to be 
favored over NS-SXTs. Of the known NS-SXTs,  Aql X-1 has been found in recent 
RXTE observations (e.g. White \& Zhang 1997) to have a $\sim 500Hz$ pulsation, 
which could put Aql X-1 as a promising MSP progenitor. However, since it has 
a $\sim 1yr$ recurrence time scale and $\sim$1 month outburst 
duration and so is more similar to the quasi-persistent source 4U1608-52 and 
the atoll sources, for which the birthrate problem exists, it may not represent 
the required class of  NS-SXT progenitor. In fact only one bona-fide NS-SXT 
system is known: Cen X-4, which shows deep quiescence between strong 
outbursts every $\sim$ 10-20 years. More detections of NS-SXTs 
(as discussed below) are definitely required for 
them to be regarded as likely MSP progenitors.

Recently, Gotthelf and Kulkarni (1997) have reported the detection of an
unusual burst source in the globular cluster M28. They attributed the  
unusually low burst luminosity to the low mass accretion rate onto a 
magnetized neutron star. The 
possible type I burst could occur 
in the magnetized accretion column which naturally increases the
effective mass accretion rate and density on the stellar surface. 
Even if such sources eventually show up in great numbers 
as a potentially important new class of NS-SXTs,  
the deduced magnetic field strength (based on the non-detection of pulsations 
and hence extremely slow spin) is much too large to be compatible 
with the typical field strength needed for MSP progenitors. 

\section{Summary and Discussion}

We have found that low luminosity LMXBs are hard to spin-up to MSPs unless they
have magnetic fields substantially lower 
($\simle 5 \times 10^7$ G) than those of the observed MSPs 
with B $\simge 10^8$ G. Thus the LMXB phase would have to 
be followed by a phase of field amplification to produce the currently
observed MSPs. 
Even when the magnetic fields are sufficiently weak, the long spin-up time scale
$\sim 10^{10}yr$ for low mass accretion rates $\ll 10^{15} g/s$ poses another
problem. If, as is likely for ADAF 
models (Narayan \& Yi 1995), the accretion flow becomes 
quasi-spherical and sub-Keplerian for ${\dot M}< 10^{16} g/s$, 
it could rule out {\it any} low luminosity population
and potentially a significant fraction of atoll sources as a possible
source of the MSPs. 
Thus we conclude that the ``standard 
model" for the formation of MSPs from LMXBs requires that they form from a 
larger population of normally quiescent 
SXT systems containing NSs with B $\simge 10^8$ G. If these sources have 
fractional duty cycle f = 0.01 
(i.e. accrete at $\sim10^{18}$ g/s for $\sim$1 month every $\sim$10 yr), 
they could provide the observed MSP population if their parent population is 
$\sim10\times$ the observed persistent atoll source population (since, 
although only $\sim 10^7$ yr is needed for spin-up, the $\sim 1$\% duty cycle 
means that an effective lifetime needed is 10$^9$ years, or the same as that 
for the persistent sources). 
More generally, a population of quiescent NS-SXTs with peak 
accretion rates ${\dot M}_{18}$ (in units of $10^{18}$ g/s) 
and on-time duty cycle f$_{-2}$ (in units of 0.01) could 
only solve the MSP birthrate problem if at any given time there 
are N$_{NS-SXT} \sim$ 10/(${\dot M}_{18}$f$_{-2}$) yr$^{-1}$ 
in outburst. Thus there would need be $\sim$10 NS-SXTs 
like Cen X-4 (with ${\dot M}_{18}$ $\sim$ 1 and  f$_{-2}$ $\sim$ 1) 
or $\sim$100 like Aql X-1 (with ${\dot M}_{17}$ $\sim$ 1 
and  f$_{-2}$ $\sim$ 10) in their ``on'' state at any given 
time, which is clearly not consistent with observations. 
Although the discovery of SAXJ1808.4-3658 may point to a 
larger population of Aql X-1 type NS-SXTs, 
the apparent rate of these new bursters discovered by RXTE or 
BeppoSAX thus far appears much too low: perhaps at most 
$\sim$3 ``new'' (recurrent) 
bursters that are ``on'' at any given time. 
A recent example is the RXTE re-discovery of 
the burster XTEJ1806-246 (Marshall and Strohmayer 1998),  
which is very likely the burster 2S1803-245 originally 
detected with SAS-3. 
Thus the required  large population of 
NS-SXTs are simply not seen (it would be 
comparable in number to the known number of persistent atoll 
sources), and so we conclude 
there is as yet no compelling evidence for MSP formation from NS-SXTs. 

A new class of MSPs could possibly form from an as yet undetected population 
of  persistent, low luminosity LMXBs. These would produce a low magnetic 
field ($<10^7G$), radio weak MSPs with typical radio luminosities 
$\sim 10^{28} erg/s$ if the radio efficiency is $\sim 1\%$. 
If the magnetic field is much weaker than $\sim 10^7G$, the MSPs would reside 
in the region below the pulsar death line, which makes the detection of these 
pulsars virtually impossible. If a significant number of low luminosity LMXBs
spin-up to MSPs, they could appear as a group of radio-weak MSPs distinct
from the known MSPs.

Given the difficulties of the ``standard model'', even for NS-SXTs, we
consider finally the accretion-induced collapse (AIC) model 
for ``direct'' production of MSPs without the LMXB progenitor phase 
(Michel 1987, Grindlay \& Bailyn 1988, Bailyn \& Grindlay 1990).
In the AIC scenario, white dwarfs collapse to neutron stars after reaching 
the critical Chandrasekhar limit $\sim 1.4M_{\odot}$ through accretion. 
It is believed that AIC occurs for high mass accretion rates
$\simge 2.5\times 10^{18} g/s$ (e.g. Nomoto \& Kondo 1991, 
Livio \& Truran 1992). The pre-AIC magnetic field on the WD will play an 
important (and poorly determined) role in the formation of MSPs. 
For magnetic WDs ($B\simge10^6$ gauss)  and initially slowly rotating white 
dwarfs accreting at a rate in the range $\sim 1\times 10^{18}-7\times 10^{18} 
g/s$ and assuming that the magnetic fields are flux-frozen and amplified by 
compression during collapse,
Yi \& Blackman (1997) have found that the magnetic field of AIC-produced 
pulsars would be correlated with the 
pulsar spin as $\Omega_{*,4}\sim \eta B_{*,11}^{-4/5}$ (cf. Narayan \& Popham
1989) where $\Omega_{*,4}=\Omega_*/10^4s^{-1}$, $B_{*,11}=B_*/10^{11}$G, and
$\eta$ is a constant of order unity. 
In Figure 1b, the dotted line is the hypothetical birth 
line of strongly magnetized pulsars from AIC. This correlation indicates that
the field strength of pulsars with $\Omega_*\sim 10^3s^{-1}$ would be
typically close to $\sim 10^{13}$G, which are many orders of magnitude larger 
than the observed MSPs. Unless these fields decay rapidly, or if magnetic 
flux is not conserved during collapse in AIC (e.g. if instead field is 
annihilated), the AIC-produced MSPs would be very bright and lose their 
rotational kinetic energy at a rate $L_{*,EM}\sim 10^{44} \Omega_{*,4}^{3/2} 
erg/s$ with the characteristic emission time scale 
$\sim I_*\Omega_*^2/2L_{*,EM}\sim 2\times 10^8\Omega_{*,4}^{1/2}s$.
Such a short time scale indicates that the pulsars could evolve almost
instantaneously to long $P_*$'s.
They could be an abundant source of LMXBs if accretion resumes
after $\sim 10yr$ of intense electromagnetic activity.
These AIC-produced LMXB's would produce MSPs similar to the observed ones
only when the magnetic field somehow decays by several orders of magnitude
during the LMXB phase. 
However, by requiring a subsequent LMXB phase, they obviously do not
solve the observed LMXB vs. MSP birthrate problem and so we do not 
consider this further (for MSP production). 

For the case of weakly magnetized ($\simle 10^{4-5}G$) white dwarfs, 
AIC could directly produce low-field MSPs but only if the collapse itself 
is not halted by rotation. To produce the observed $\sim$2 ms spin periods, 
the pre-AIC WD must have spin period $\sim$20 s (assuming conservation of 
angular momentum in the collapse from a 1.4M$_ {\odot}$ WD of radius 
$\sim 10^3$ km to a 10 km radius NS). However, for low magnetic field, 
these stars are not braked by the magnetic fields and so the spin period of 
the pre-collapse WD could instead be spun up to the minimum  
($(4\pi^2R_{wd}^3/GM_{wd})^{1/2}\sim$1s) period where collapse would be 
halted unless the excess angular momentum is first removed. Thus for AIC to be 
completed without producing the so-called rotationally supported ``fizzlers", 
angular momentum must be efficiently removed during or before collapse 
(e.g. Tohlin 1984).  This implies that AIC would produce the MSPs only if 
white dwarfs stop spinning-up despite continued accretion. Thus the possibility
of spin-down near the break-up period, despite continuous accretion 
(Popham \& Narayan 1991, Paczynski 1991, Bisnovatyi-Kogan 1993), 
is an intriguing possible solution to the MSP birthrate problem.

We conclude that if 
an AIC process produces the bulk of the observed MSPs,  
it is necessary to shed angular momentum (low B field WD case) or magnetic 
field (high B field WD case) either prior to or during the collapse. 
Since only the former process directly produces a MSP without 
an LMXB, it is preferred. However both processes may occur and 
in fact may relate the problems of MSP formation with that of 
the origin of gamma-ray bursts, as we discuss in a forthcoming paper
(work in preparation).  
\medskip

We thank the referee, S. Kulkarni, and J. McClintock for useful
comments on the manuscript. This work was supported in part by 
NASA/LTSA grant NAG5-3256 (to JEG) and by the SUAM Foundation (to IY).

\vfill\eject

\centerline{Figure Captions}

\noindent
Figure 1a: Spin-up lines in the $B_*$ vs. $P_*$ plane. The x's are
the observed pulsars adopted from Taylor et al. (1995). 
The dashed line is the pulsar death line, $B_*\sim 2\times 10^{11} P_*^2$. 
The solid lines are the spin-up lines for ${\dot M}$,
$10^{18} g/s$, ${\dot M}_c=3.5\times 10^{16} g/s$, $10^{16} g/s$, 
$6.3\times 10^{15} g/s$, and $10^{13} g/s$ from top to bottom, respectively.
The thick solid lines are relevant for typical atoll sources.

\vskip 0.5cm

\noindent
Figure 1b: Similar to Figure 1a with the sub-Keplerian rotation
($\omega_a=0.2$) for ${\dot M}<{\dot M}_c$. The spin-up line for a given 
${\dot M}<{\dot M}_c$ moves down due to the sub-Keplerian rotation. 
>From top to bottom, ${\dot M}$'s are $3.5\times 10^{16}g/s$, $10^{16} g/s$, 
$6.3\times 10^{15}g/s$, and $10^{13}g/s$, respectively.
The solid box in the bottom left indicates the location of the "typical" 
MSPs spun-up by the sub-Keplerian accretion flows.
The dotted line is a possible birth line from AIC for 
${\dot M}=3.2\times 10^{18} g/s$. 
The thick solid lines are relevant for typical atoll sources.

\vfill\eject

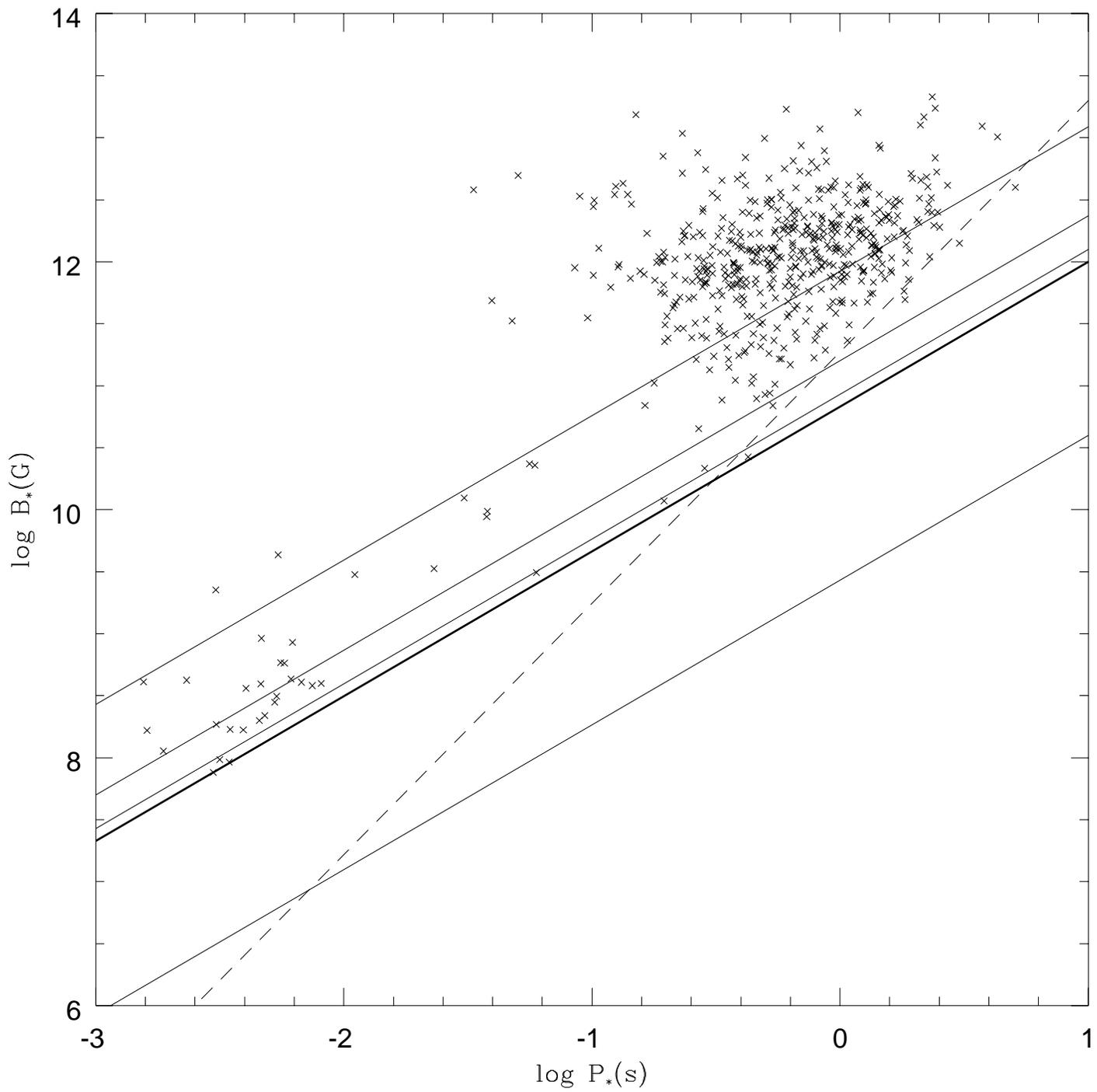
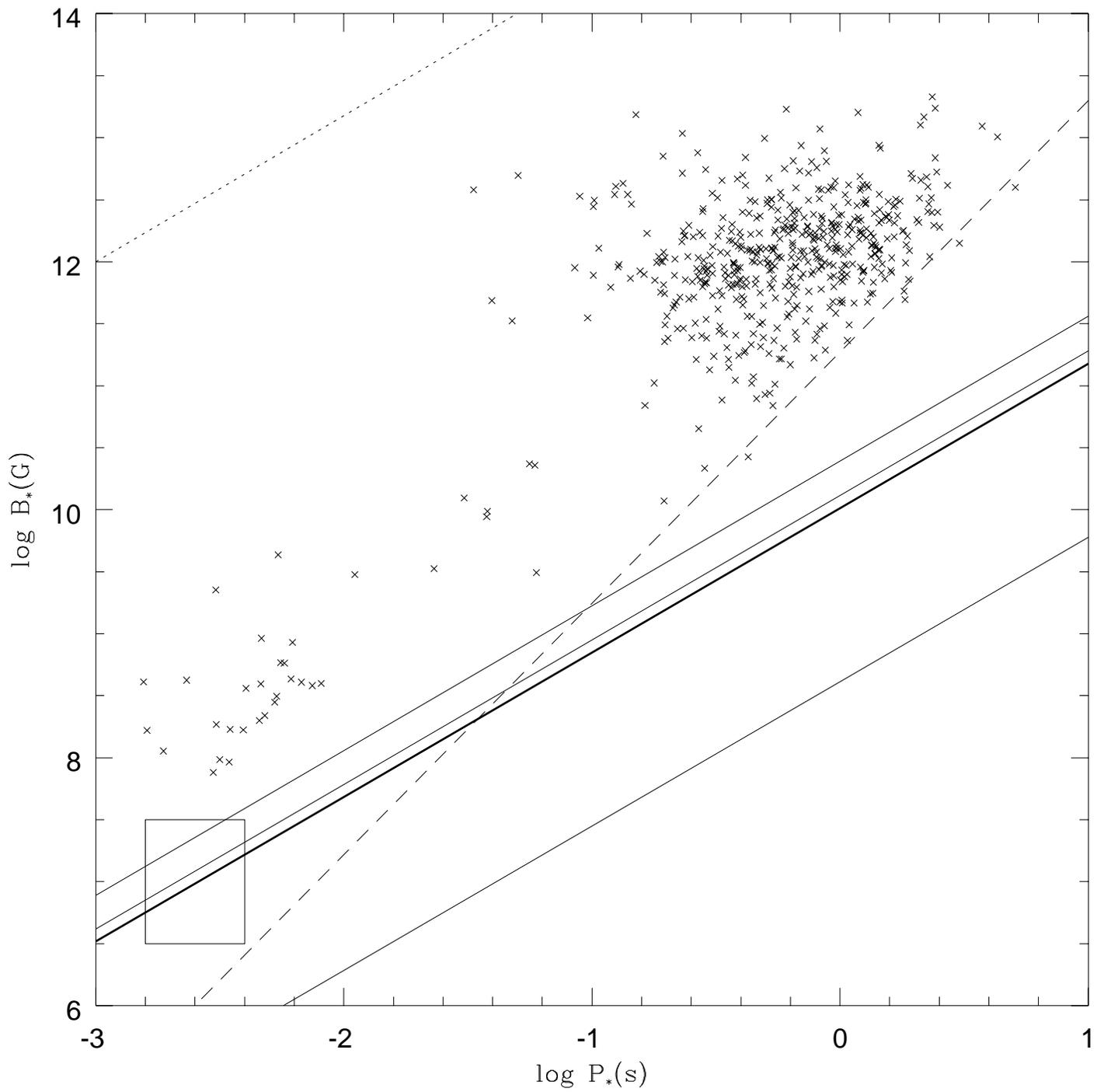
\end{document}